\documentclass[onecollarge,natbib]{svjour2}
\bibpunct{[}{]}{;}{n}{}{,} 
\smartqed  
\usepackage{slashed}
\usepackage{graphicx}
\usepackage{latexsym}
\usepackage{amsmath}
\journalname{Few-Body Systems}
\begin{document}

\title{Application of Coherent State Approach for the cancellation of Infrared divergences to all orders in LFQED 
}


\author{Jai More    \and
        Anuradha Misra 
}

\institute{Jai More \at
          Department of Physics, University of Mumbai, 
SantaCruz(East), Mumbai-400098, India. \\
\email more.physics@gmail.com
            \\
           \and
           Anuradha Misra  \at
           Department of Physics, University of Mumbai,
SantaCruz(East), Mumbai-400098, India.\\ 
\email anuradha.misra@gmail.com
}

\date{Received: date / Accepted: date}

\maketitle

\begin{abstract}
We sketch an all order proof of cancellation of infrared (IR) divergences in Light Front Quantum Electrodynamics (LFQED) using a 
coherent state formalism. In this talk,  it has been shown that the true IR divergences in fermion self energy are eliminated 
to all orders in a light-front time-ordered perturbative calculation if one uses coherent state basis instead of the usual
Fock basis to calculate the Hamiltonian matrix elements.
\end{abstract}

\section{Introduction}
\label{intro}
In the LSZ formalism,
the asymptotic states are taken as free states and the S--matrix elements are the residues of the poles
that arise in the Fourier transform of the correlation functions when four-momenta of the external
particles are put on-shell. Therefore, the initial and final states used to calculate the transition 
matrix elements are taken to be the Fock states. However, in Quantum Electrodynamics (QED) the asymptotic states are not free states and the fermion is 
accompanied by soft photons i.e. photons with very low momentum. In an actual experiment, due to the finite size of the detector, the charged particle 
can be accompanied by any number of such photons, which are
 source of Infrared (IR) divergences. In the soft photon limit, the virtual and real emission are indistinguishable. So when we are dealing with a virtual photon correction in a 
process, we need to take into account emission of an infinite number of real soft photons also. Hence, the physical state should be 
defined as a set of states with an infinite number of soft photons.

A method of asymptotic dynamics was developed by Kulish and Faddeev (KF) to address the issue of cancellation of IR divergences at amplitude level \cite{KUL70}. They were the first to show that in QED, the asymptotic Hamiltonian does not coincide with the 
free Hamiltonian. KF constructed the asymptotic Hamiltonian $V_{as} $ for QED thus modifying the asymptotic 
condition to introduce a new space of asymptotic states.

KF further modified the definition of S$-$matrix and showed that it is free of IR divergences. Thus the method of asymptotic 
dynamics proposed by KF replaces, the free Hamiltonian by an asymptotic Hamiltonian which takes into account the long 
range interaction between incoming and outgoing states and can be used to construct a set of coherent states as the asymptotic states. 
The transition matrix elements formed by using these states are then IR finite.

 In Light Front Field Theory (LFFT),  there are two kinds of IR divergences viz. the {\it true} 
IR divergences which are the bona-fide divergences of equal-time field theory and which appear when both $k^+$ and ${\bf k_\perp}$ approach zero and 
the {\it spurious} IR divergences that are just a manifestation of 
ultra-violet divergences of equal-time theory. The KF method has been used by various authors \cite{BUT78, DAH81, GRECO78, VARY99} in the context of equal time theories. 
In LFFT, a coherent state formalism was developed in Ref.~\cite{ANU94, ANU96, ANU00} as a possible method to deal with the {\it true} IR divergences 
in one loop vertex correction for LFQED and LFQCD.   It has also been shown that the IR divergences in the
fermion self-energy correction at the two-loop order cancel
in LFQED \cite{JAI12, JAI13} if one uses the coherent state approach.

In QED, a general approach to treat IR divergences was
given by Yennie et al. \cite{YENNIE61}. Following the work of Yennie et al., Chung \cite{CHU65} showed
that IR divergences indeed cancel to all orders in perturbation
theory at the level of amplitude. KF approach was applied by Greco etal\cite{GRECO78} to study the IR behavior  of  non abelian gauge theories using coherent states of definite color and factorized in  fixed angle regime. 
We shall present a general proof of all order 
cancellation of IR divergences in fermion mass renormalization in LFQED by using the coherent state method. The details of the calculation can be found in Ref.\cite{JAI14}. 
\section{Coherent State Formalism in Light Front Field Theory}\label{sec:1}
The coherent state method is based on the observation that for theories with long range interactions or theories  having bound states as asymptotic states, the total Hamiltonian does not reduce to the free field Hamiltonian in the limit $\vert t \vert \rightarrow \infty$. In LFFT, the asymptotic Hamiltonian $H_{as}$,  is evaluated by taking the limit $\vert x^+\vert\rightarrow \infty$ of the interaction Hamiltonian. Each term in the interaction Hamiltonian 
$H_{int}$ has a light-cone time dependence of the form  $\exp[-i(p_1^-+p_2^-+\cdots+p_n^-)x^+]$ and therefore, if  $(p_1^-+p_2^-+\cdots+p_n^-)$ vanishes at  some vertex, then the corresponding term in $H_{int}$ does not vanish in large $x^+$ limit. Thus, the total Hamiltonian can be written as 
\begin{equation}
H = H_{as} + H_I^\prime
\end{equation} 
where 
\begin{equation}
H_{as}(x^+) = H_0 +V_{as}(x^+)
\end{equation}
In the Schr\"odinger representation, light cone time evolution operator $U_{as}(x^+) $  satisfies 
\begin{equation} 
i\frac{dU_{as}(x^+)}{dx^+} = H_{as}(x^+)U_{as}(x^+)
\end{equation}
and can then be used to generate an asymptotic space
\begin{equation}
{\cal H}_{as} = \exp[-\Omega^A(x^+)]{\cal H}_F
\end{equation}
from the usual Fock space ${\cal H}_F$, in the limit $x^+ \rightarrow -\infty$, where
$\Omega^A(x^+)$ is the asymptotic evolution operator defined by 
\begin{equation} 
U_{as}(x^+) = \exp[-iH_0x^+] \exp[\Omega^A(x^+)]
\end{equation}
The coherent states is defined as
\begin{equation}
\vert n : coh \rangle = \exp[-\Omega^A] \vert n \rangle
\end{equation} 
KF proposed the method of asymptotic dynamics \cite{KUL70} in the context of equal time QED wherein the asymptotic Hamiltonian is not just the free Hamiltonian but also 
contains the large time limit of those terms in
interaction Hamiltonian which do not vanish at infinitely large times. This Hamiltonian is  then be used to construct the asymptotic M\"oller operator and the coherent states. 
The true IR divergences corresponding to $k^+,~{\bf k}_\perp \rightarrow 0$ are expected to disappear when one uses this 
coherent state basis to calculate the transition matrix elements. 
The light-cone  time dependent interaction  Hamiltonian is given by
\begin{displaymath}
H_I(x^+)=V_1(x^+)+V_2(x^+)+V_3(x^+)
\end{displaymath}
where \cite{ANU94}
\begin{align}\label{V_1(x^+)}
V_1(x^+)= e\sum_{i=1}^4 \int d\nu_i^{(1)}[ e^{-i \nu_i^{(1)} x^+} {\tilde h}_i^{(1)}(\nu_i^{(1)})
+ e^{i \nu_i^{(1)} x^+}
{\tilde h}^{(1)\dagger}_i (\nu_i^{(1)})]
\end{align}
${\tilde h}^{(1)}_i(\nu^{(1)}_i)$ are three point QED interaction vertices. 
  At asymptotic limits, non-zero contributions to $V_1(x^+)$ come from regions
where $\nu_i^{(1)} \rightarrow 0$. It is easy to see that $\nu_2^{(1)}$ and $\nu_3^{(1)}$ are always non-zero and therefore, 
$\tilde h_2$ and $\tilde h_3$ do not appear in the asymptotic Hamiltonian. Thus, the 3$-$point asymptotic Hamiltonian is 
defined by the following expression \cite{ANU94}
\begin{eqnarray}\label{V_1as}
V_{1as}(x^+) = e \sum_{i=1,4} \int d\nu_i^{(1)} \Theta_\Delta(k)
 [ e^{-i \nu_i^{(1)} x^+} \tilde h_i^{(1)}(\nu_i^{(1)})
+ e^{i \nu_i^{(1)} x^+}
\tilde h^ {\dagger}_i (\nu_i^{(1)})] \;
\end{eqnarray}
where $\Theta_{\Delta}(k)$ is a function which takes value 1 in the asymptotic region and is zero elsewhere. The detailed discussion of construction of coherent state basis can be found in  Ref.~\cite{ANU94, JAI12, JAI13}.
The asymptotic states can be defined in the usual manner by 
\begin{equation}
\vert n \colon coh \rangle = \Omega_{\pm}^A \vert{n}\rangle    \;,
\end{equation}
where $\vert{n}\rangle$ is a Fock state and $\Omega_{\pm}^A$, the asymptotic M\"oller operator is defined by
\begin{equation}
\Omega_{\pm}^A = T~\exp\biggl[ -i \int^0_{\mp \infty}[ V_{1as}(x^+)+ V_{2as}(x^+)]dx^+ \biggr] \;
\end{equation}
where $ V_{2as}(x^+)$ is the asymptotic limit of the four-point instantaneous interaction terms of LFQED Hamiltonian. 
The strategy to construct an all order proof of
the cancellation of IR divergences in the fermion self energy
correction using the coherent state basis is based on  the method of induction\cite{JAI14}.
\section{All order proof of cancellation of Infrared divergences in LFQED}\label{sec:2}
The all order proof of cancellation of IR divergences in LFQED is based on the method of induction \cite{YENNIE61} and involves the 
following steps:
\begin{itemize}
\item[I] We first show the cancellation of IR divergences up to $O(e^4)$ in LFQED \cite{JAI12, JAI13} in our new graphical notation.
\vskip 0.25cm
\item[II] We assume that IR divergences cancel up to $O(e^{2n})$. We express this IR finite $O(e^{2n})$ contribution graphically as in Fig.1.  \vskip 0.25cm
\item[III] Then we express the $O(e^{2(n+1)})$ contribution in terms of IR finite $O(e^{2n})$ matrix elements by adding virtual photon lines for Fock state diagrams 
and real photon lines for coherent state diagrams\vskip 0.25cm 
\item[IV] Finally we show that these additions to $O(e^{2n})$ diagrams conspire to cancel the  IR divergences up to $O(e^{2(n+1)})$ between the real and virtual diagrams.
\end{itemize}
\begin{figure}[h]
\centering
 \includegraphics[scale=1.2]{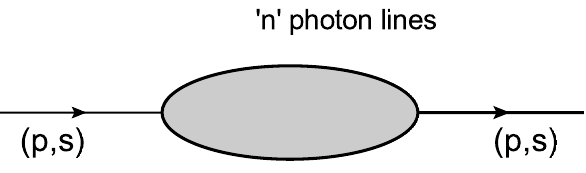}
 \caption{IR finite blob representing the sum of all $O(e^{2n})$  diagrams in coherent state basis}
 \label{blob1}
\end{figure}
We give below a brief description of the proof. The details of the proof are in Ref.~\cite{JAI14}. 
The general expression for transition matrix element in $O(e^{2n})$ is a sum of terms of the form:
\begin{align}
 T^{(n)}_j&=-\frac{e^{2n}}{2p^+ (2\pi)^{3n}}\int \prod_{i=1}^n\frac{d^3k_i}{2k_i^+ 2p_{2i-1}^+}\\
 &\times\frac{\overline{u}(\overline{p},s_1)\epsilon\llap/_1(\not p_1+m)\epsilon\llap/_2(\not p_2+m)\cdots \, \cdots \cdots(\not p_i+m)\epsilon\llap/_i u(p, s_i)}
 {\displaystyle \prod_r (p^--p_r^--\sum_i k_i)} 
\end{align} 
We express the total $O(e^{2n})$ contribution as 
\begin{equation}
T^{(n)}=\sum_j T^{(n)}_j=\sum_j \frac{\overline{u}(\overline{p},s^\prime)\mathcal{M}_n^{(j)} u(p,s)}{\mathcal{D}^{(j)}}
\end{equation}
where $j$ is summed over all possible diagram in  $O(e^{2n})$  and will be assumed to be IR divergence free.\\
\begin{equation}
\mathcal{D}^{(j)}=\displaystyle \prod_{i=1}^n \mathcal{D}_i^{(j)} 
\end{equation}
$\mathcal{M}_n^{(j)}$ is represented graphically by the blob in Fig 1. 
We have already given the proof of cancellation of of IR divergences in $\delta m^2$ up to $O(e^4)$ in Ref.~\cite{JAI12}. Now, we revisit this proof in the new graphical notation to make our strategy more lucid. Consider the $O(e^4)$ corrections which are
represented by the two diagrams on the rhs of Fig. \ref{Fig3Oe4}. It has
been shown in Refs. \cite{JAI12, JAI13} that the sum of these two is IR
finite. \vskip -0.5cm
\begin{figure}[h]
\centering
 \includegraphics[scale=1.0]{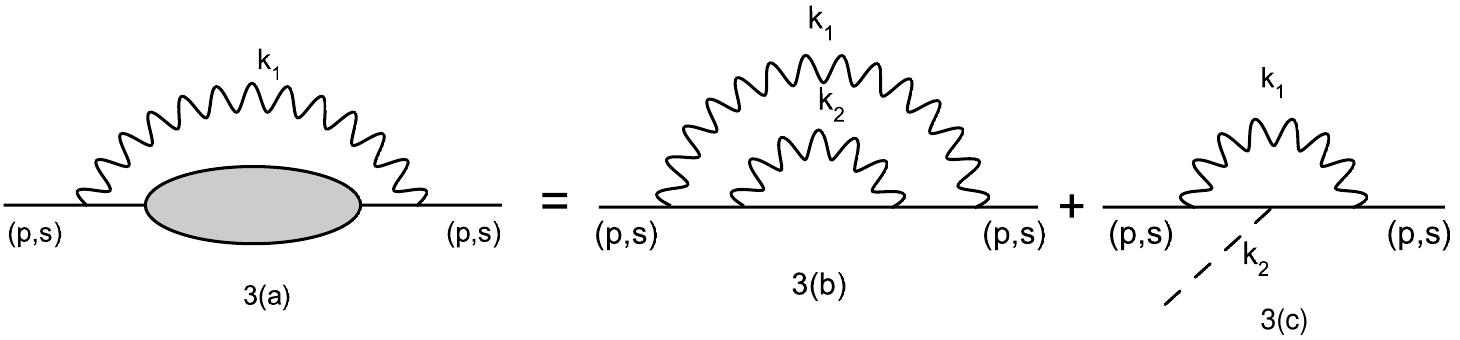}
\caption{IR finite $O(e^2)$ blob with an external photon line results in a sum of $O(e^4)$ diagrams}
\label{Fig3Oe4}
\end{figure}
In our new notation  
\begin{align}\label{T3}
T_{3a}^{(2)}&=T_{3b}^{(2)}+T_{3c}^{(2)}\nonumber \\
	    &=\frac{e^2}{(2\pi)^3}\int \frac{d^3k_1}{2k_1^+}\frac{\overline{u}(p, s)\epsilon\llap/(k_1)(\slashed p_1+m)\mathcal{M}_2^{(j)} (\slashed p_1+m) \epsilon\llap/(k_1) u(p,s)}
	      {(p\cdot k_1)^2 \mathcal{D}^{(j)}}
 \end{align}
where $\mathcal{M}_2^{(j)}$ represents the IR finite blob of Fig.1 without the external lines. \vskip -0.5cm
\begin{figure}[h]
\centering
 \includegraphics[scale=1.2]{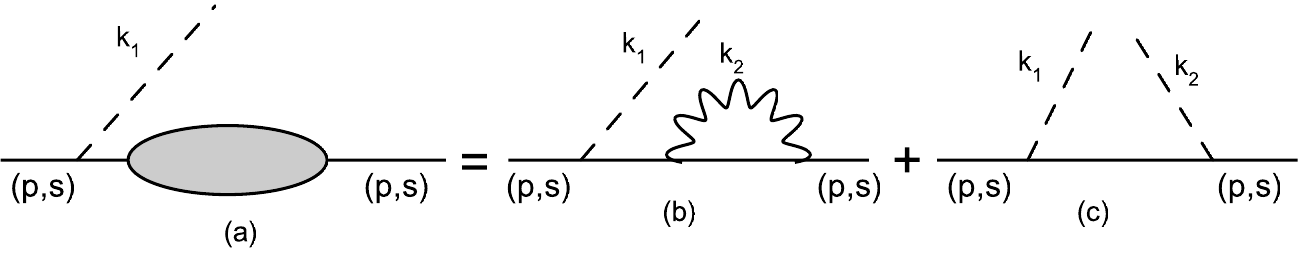}
\caption{IR finite $O(e^2)$ blob with an external photon line results in a sum of $O(e^4)$ diagrams in coherent state basis}
\label{Fig4cohOe4}
\end{figure} \vskip -0.25cm
Fig.~\ref{Fig4cohOe4} represents the additional  $O(e^2)$ diagrams in coherent state basis  which cancel IR divergences in Fig. \ref{Fig3Oe4}. The additional contribution due to these in our new notation is 
\begin{align}\label{T4}
T_{4a}^{(2)}=-\frac{e^2}{(2\pi)^3}\int \frac{d^3k_1}{2k_1^+}\frac{\overline{u}(p, s)\mathcal{M}_2^{(j)}(\slashed p_1+m)\epsilon\llap/(k_1) u(p,s)(p\cdot k_1)}
 {(p\cdot k_1)^2 \mathcal{D}^{(j)}}
 \end{align}
In the limit $k^+ \rightarrow 0, k_\perp \rightarrow 0$ 
the sum of the terms on rhs of Eqs.~(\ref{T3}) and (\ref{T4}) is IR finite. A similar argument can be constructed for the
remaining two diagrams which are graphically representated by Fig.~\ref{Fig5Oe4}. 
\begin{figure}[h]
\centering
\includegraphics[scale=1]{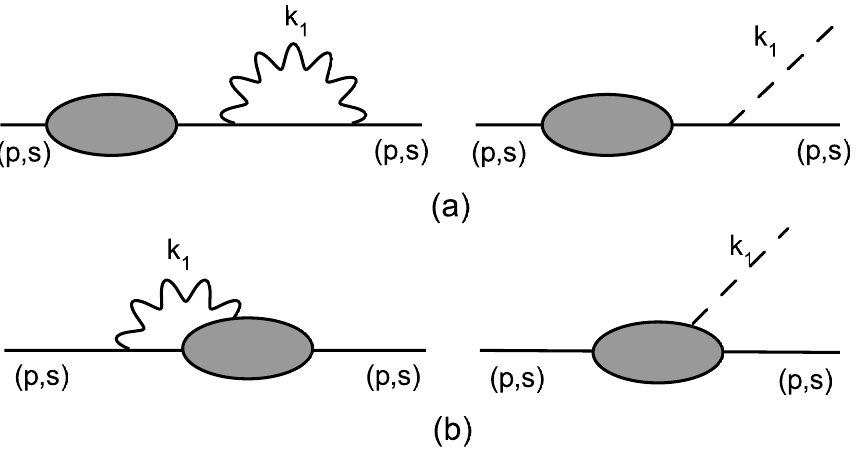}
\caption{Additional diagrams corresponding to $O(e^4)$ contributions in Fock and coherent state bases.}
\label{Fig5Oe4}
\end{figure} 

\begin{figure}[h]
\centering
\includegraphics[width=24cm,height=6cm,keepaspectratio]{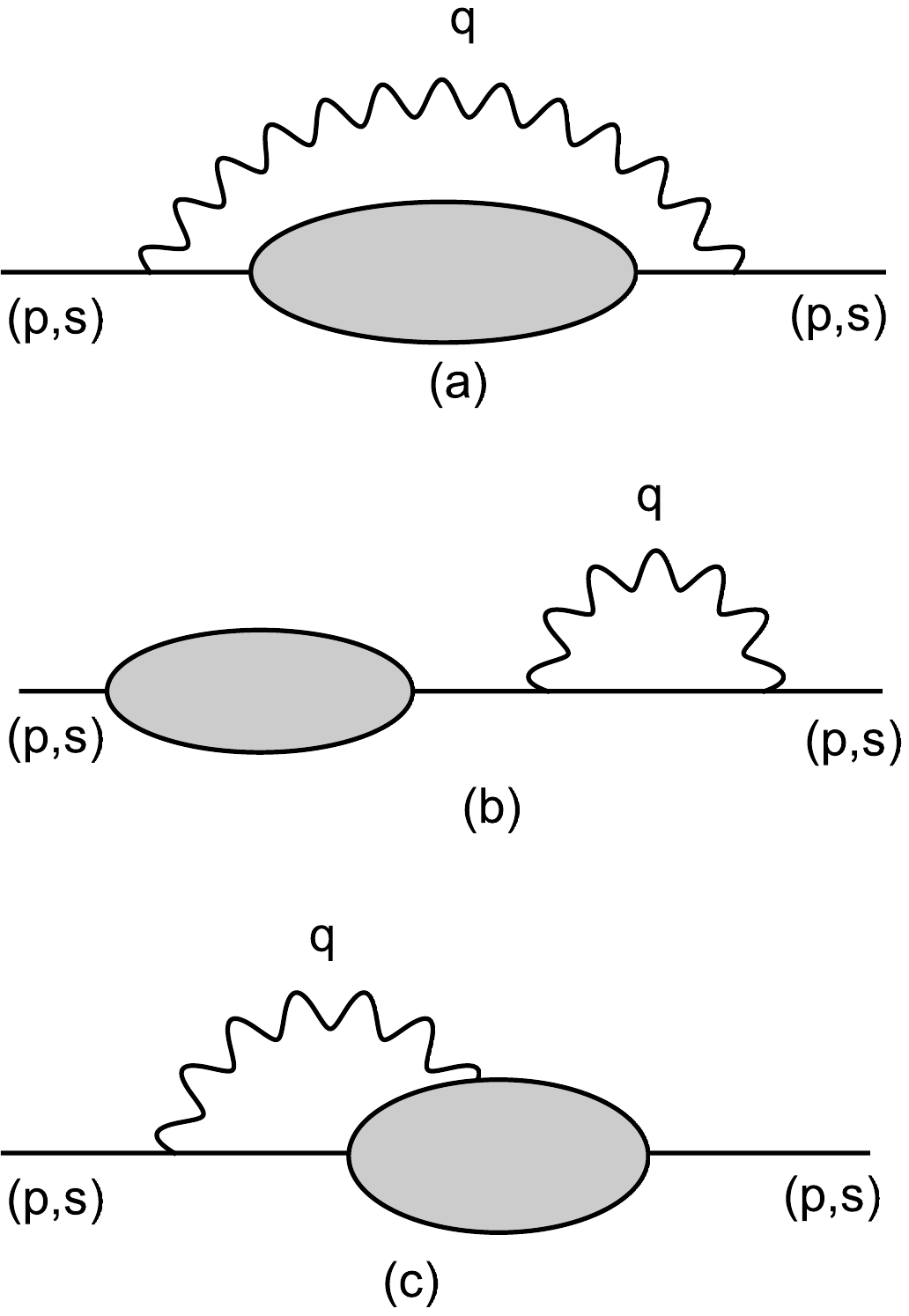}
 \caption{Addition of a photon line to $O(e^{2n})$ blob in Fock basis}
\label{blobn2}
\end{figure}

Now, we consider an $O(e^{2n})$ blob which we will assume
to be free of IR divergences as shown in Fig. \ref{blob1}. We shall show that the
cancellation of IR divergences in the $O(e^{2(n+1)})$ contribution
to fermion mass renormalization in LFQED follows form the assumption that  $O(e^{2n})$ diagrams are free of IR divergences. 
To construct an $O(e^{2(n+1)})$ diagram in Fock basis, we can add a photon to $n^{th}$ order blob in three different ways as shown in Fig.~\ref{blobn2}.

\begin{figure}[h]
\centering
\includegraphics[width=24cm,height=6.5cm,keepaspectratio]{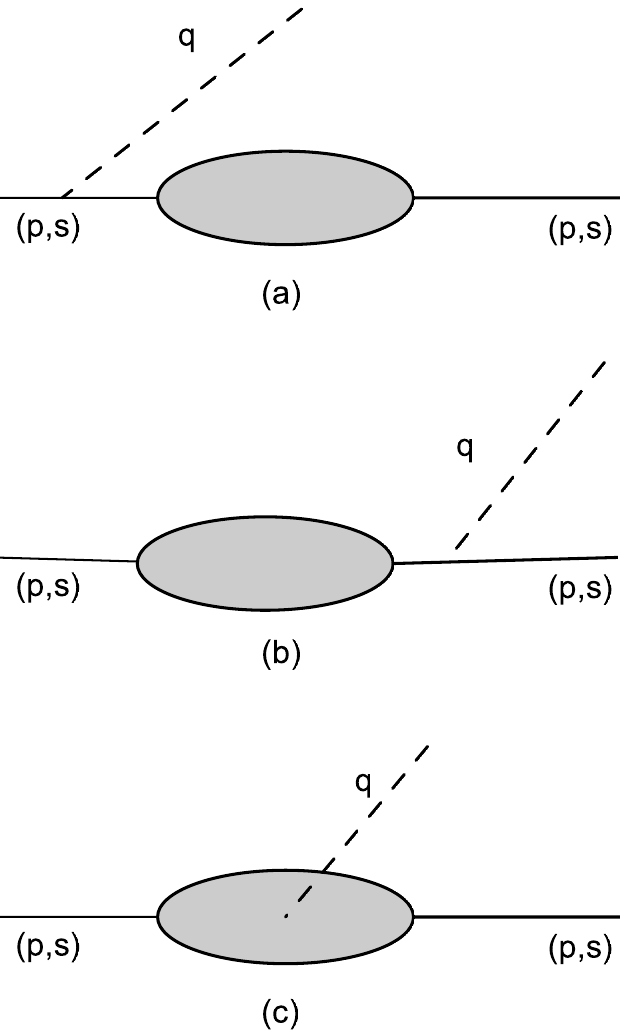}
\caption{Addition of a photon line to $O(e^{2n})$ blob in coherent state basis}
\label{blob3}
\end{figure}

The contributions coming from the diagram in Figs.~\ref{blobn2}(a) , (b) and (c), in the limit $q^+ \rightarrow 0, {\bf q_\perp} \rightarrow 0$   are given by 
\begin{align}\label{T6a}
 T^{(n+1)}_{5a}=&\frac{e^2}{(2\pi)^3}\int \frac{d^3q}{2q^+}\frac{\overline{u}(p,s)\epsilon\llap/(q)(\slashed P+m)\mathcal{M}_n^{(j)} (\slashed P+m) \epsilon\llap/(q) u(p,s)}
 {(p\cdot q)^2 \mathcal{D}^{(j)}}\\ \label{T6b}
 T^{(n+1)}_{5b}=&-\frac{e^2}{(2\pi)^3}\int \frac{d^3q}{2q^+}\frac{\overline{u}(p,s)\epsilon\llap/(q)(\slashed P+m)\epsilon\llap/(q) (\slashed p^\prime+m) \mathcal{M}_n^{(j)} u(p,s)}
 {(p\cdot q)(p^--p^{\prime-})\mathcal{D}^{(j)}}  \\ \label{T6c}
  T^{(n+1)}_{5c}=&\frac{e^2}{(2\pi)^3}\int \frac{d^3q}{2q^+}\frac{\overline{u}(p,s)\mathcal{M}_n^{(j)} (\slashed P+m)\epsilon\llap/(q) u(p,s)}
  {(p\cdot q)\mathcal{D}^{(j)}}
\end{align}
where $P=p-q$ and  $p^\prime = p$.\\
There are  additional contributions in $(n+1)^{th}$ order, when we use the coherent state basis which are  shown  in Fig.~\ref{blob3} and are  given by
\begin{align}\label{T7a}
 T^{\prime(n+1)}_{6a}=&-\frac{e^2}{(2\pi)^3}\int \frac{d^3q}{2q^+}\frac{\overline{u}(p,s)\mathcal{M}_n^{(j)}(\slashed P+m)\epsilon\llap/(q) u(p,s)(p \cdot \epsilon(q))\,\Theta_\Delta(q)}
 {(p\cdot q)^2\mathcal{D}^{(j)}}\\ \label{T7b} 
 T^{\prime(n+1)}_{6b}=&\frac{e^2}{(2\pi)^3}\int \frac{d^3q}{2q^+}\frac{\overline{u}(p,s)\epsilon\llap/(q) (\slashed p^\prime+m) \mathcal{M}_n^{(j)} u(p,s)(p\cdot \epsilon(q))\,\Theta_\Delta(q)}
 {(p\cdot q)(p^--p^{\prime-})\mathcal{D}^{(j)}}\\  \label{T7c}
  T^{\prime(n+1)}_{6c}=&-\frac{e^2}{(2\pi)^3}\int \frac{d^3q}{2q^+}\frac{\overline{u}(p,s)\mathcal{M}_n^{(j)} u(p,s)(p \cdot \epsilon(q))\,\Theta_\Delta(q)}{(p\cdot q)\mathcal{D}^{(j)}}
\end{align}
In the limit, $q^+\rightarrow 0$, ${\bf q}_\perp \rightarrow 0$, $\slashed P\epsilon\llap/(q) \rightarrow p \cdot \epsilon$. 
As a result, the IR divergences sum of Eqs.~(\ref{T6a}) -- (\ref{T6c}) are exactly cancelled by the sum of Eqs.~(\ref{T7a}) -- (\ref{T7c}) 
leading to   an infrared finite result.
\section{Conclusions}\label{sec:3}
We have demonstrated,  using the  coherent state formalism,   that the true IR divergences in self energy correction cancel to all order in 
LFQED. 
%
%

\begin{acknowledgements}
J.M. would like to thank ILCAC for honouring with Mc Cartor Travel Grants for attending LC2014 and would also like to thank the LC 2014 organizers for their kind hospitality. 
AM would like to thank organizers of LC2014 and Department of Physics, NCSU for their warm hospitality and DAE-BRNS for 
Grant No. 2010/37P/47/BRNS under which the work was done.
\end{acknowledgements}



\end{document}